\input harvmac.tex
\let\includefigures=\iftrue
\newfam\black
\includefigures
\input epsf
\def\figin{\epsfcheck\figin}\def\figins{\epsfcheck\figins}
\def\epsfcheck{\ifx\epsfbox\UnDeFiNeD
\message{(NO epsf.tex, FIGURES WILL BE IGNORED)}
\gdef\figin##1{\vskip2in}\gdef\figins##1{\hskip.5in}
\else\message{(FIGURES WILL BE INCLUDED)}%
\gdef\figin##1{##1}\gdef\figins##1{##1}\fi}
\def\DefWarn#1{}
\def\figinsert{\goodbreak\midinsert}
\def\ifig#1#2#3{\DefWarn#1\xdef#1{fig.~\the\figno}
\writedef{#1\leftbracket fig.\noexpand~\the\figno}%
\figinsert\figin{\centerline{#3}}\medskip\centerline{\vbox{\baselineskip12pt
\advance\hsize by -1truein\noindent\footnotefont{\bf Fig.~\the\figno:} #2}}
\bigskip\endinsert\global\advance\figno by1}
\else
\def\ifig#1#2#3{\xdef#1{fig.~\the\figno}
\writedef{#1\leftbracket fig.\noexpand~\the\figno}%
\global\advance\figno by1}
\fi
\def\Title#1#2{\rightline{#1}
\ifx\answ\bigans\nopagenumbers\pageno0\vskip1in%
\baselineskip 15pt plus 1pt minus 1pt
\else
\def\listrefs{\footatend\vskip 1in\immediate\closeout\rfile\writestoppt
\baselineskip=14pt\centerline{{\bf References}}\bigskip{\frenchspacing%
\parindent=20pt\escapechar=` \input
refs.tmp\vfill\eject}\nonfrenchspacing}
\pageno1\vskip.8in\fi \centerline{\titlefont #2}\vskip .5in}

\ifx\answ\bigans\def\tcbreak#1{}\else\def\tcbreak#1{\cr&{#1}}\fi
%

\def\scrn{{\cal{N}}}
\nref\BFSS{ T. Banks, W. Fischler, S.H. Shenker, L.Susskind,
``M-Theory As a Matrix Model: A Conjecture'', Phys. Rev. D55 (1997)
5112, hep-th/9610043.}
\nref\Motl{ L. Motl, ``Proposals on nonperturbative superstring
interactions'', hep-th/9701025.}
\nref\BS{T. Banks, N. Seiberg, ``Strings from Matrices''.
hep-th/9702187.}
\nref\DVV{R. Dijkgraaf, E. Verlinde, H. Verlinde, ``Matrix
String Theory'', hep-th/9703030.}
\nref\BD{M. Berkooz, M.R. Douglas, ``Five-Branes in Matrix
Theory'', Phys. Lett. B395 (1997) 196, hep-th/9610236.}
\nref\BSS{T. Banks, N. Seiberg, S. Shenker, ``Branes from
Matrices'', Nucl. Phys. B490 (1997) 91, hep-th/9612157.}
\nref\scatt{M. Berkooz, O. Aharony, ``Membrane Dynamics in
Matrix Theory'', Nucl. Phys. B491 (1997) 184, hep-th/9611215,\parskip=0pt
\item{}G. Lifschytz, S.D. Mathur, ``Supersymetry and Membrane Interactions in
Matrix Theory'', hep-th/9612087,
\item{}G. Lifschytz, ``Four-Brane and Six-Brane Interaction in Matrix
Theory'', hep-th/9612223,
\item{}V. Balasubramanian, F. Larsen, ``Relativistic Brane Scattering'',
hep-th/9703039,
\item{}D. Berenstein, R. Corrado, ``Matrix-Theory in Various Dimensions``, 
hep-th/9702108.}
\nref\GRT{O.J. Ganor, S. Ramgoolan, W. Taylor, ``Branes, Fluxes
and Duality in Matrix-Theory'',hep-th/9611202,\parskip=0pt
\item{}L. Susskind, ``T Duality in Matrix Theory and S Duality in Field
Theory'', hep-th/9611164,
\item{}W. Taylor, ``D-brane field theory on compact spaces'',
Phys. Lett. B394 (1997) 283, hep-th/9611042.}
\nref\D{ M.R. Douglas, ``Enhanced Gauge Symmetry in Matrix
Theory'', hep-th/9612126.}
\nref\FHR{W. Fischler, E. Halyo, A. Rajaraman, L. Susskind,
``The Incredible Shrinking Torus'', hep-th/9703102.}
\nref\FR{W. Fischler, A. Rajaraman, ``Matrix String Theory On
K3'', hep-th/9704123.}
\nref\KR{ N. Kim, S.-J. Rey, ``Matrix Theory on $T^5/Z_2$
Orbifold and Five-Brane'', hep-th/9705132,\parskip=0pt
\item{}N. Kim, S.-J. Rey, ``Matrix Theory on an Orbifold and Twisted
Membrane'', hep-th/9701139.}
\nref\FS{A. Fayyazuddin, D.J. Smith, ``A note on $T^5/Z_2$
compactification of M-theory matrix model'', hep-th/970320.}
\nref\BRS{ M. Berkooz, M. Rozali, N. Seiberg, ``Matrix Theory
Description of M-theory on $T^4$ and $T^5$'', hep-th/9704089.}
\nref\Seib{ N. Seiberg, ``New Theories in Six Dimensions and Matrix
Description of M-theory on $T^5$ and $T^5/Z_2$, hep-th 9705221}
\nref\gov{S. Govindarajan, ``A note on M(atrix) theory in seven
dimensions with eight supercharges'', hep-th/9705113.}
\nref\BR{ M. Berkooz, M. Rozali, ``String Dualities from Matrix
Theory'', hep-th/9705175.}
\nref\Wa{K. Dasgupta, S. Mukhi, ``Orbifolds of M-theory'', 
Nucl. Phys. B465 (1996) 399, hep-th/9512196,\parskip=0pt
\item{}E. Witten, ``Five-branes And M-Theory On An Orbifold'',
Nucl. Phys. B463 (1996) 383, hep-th/9512219.}
\nref\VO{ H. Ooguri, C. Vafa, ``Two-Dimensional Black Hole and
Singularities of CY Manifolds'', Nucl. Phys. B463 (1996) 55, hep-th/9511164.}
\nref\IS{K. Intriligator, N. Seiberg, ``Mirror Symmetry in Three
Dimensional Gauge Theories'', Phys. Lett. B387 (1996) 513,
hep-th/9607207.}
\nref\OO{J. de Boer, K. Hori, H. Ooguri, Y. Oz, Z. Yin, ``Mirror
Symmetry in Three-Dimensional Gauge Theories, $SL(2,Z)$ and D-Brane
Moduli Spaces'', hep-th/9612131,\parskip=0pt
\item{}J. de Boer, K. Hori, H. Ooguri, Y. Oz, ''Mirror
Symmetry in Three-Dimensional Gauge Theories, Quivers and
D-branes'',hep-th/9611063.} 
\nref\HW{A. Hanany, E. Witten, ``Type IIB Superstrings, BPS
Monopoles, And Three-Dimensional Gauge Dynamics'', hep-th/9611230.}
\nref\PZ{ M. Porrati, A. Zaffaroni, ``M-Theory Origin of Mirror
Symmetry in Three Dimensional Gauge Theories'', Nucl. Phys. B490
(1997) 107, hep-th/9611201.}
\nref\Wc{ E. Witten, ``String Theory In Various Dimensions'',
Nucl. Phys. B443 (1995) 85, hep-th/9503124. }
\nref\Ad{ P.S. Aspinwall, D.R. Morrison, `` U-Duality and
Integral Structures'', Phys. Lett. B355 (1995) 141, hep-th/9507012.}
\nref\Aa{ P.S. Aspinwall, ``K3 Surfaces and String Duality'',
het-th/9611137.}
\nref\Wd{E. Witten, ``Some Comments On String Dynamics'', 
hep-th/9507121.}
\nref\HD{ J. Distler, A. Hanany,''$(0,2)$ Noncritical Strings in
Six Dimensions'', Nucl. Phys. B490 (1997) 75, hep-th/9611104.}
\nref\Ab{ P.S. Aspinwall, D.R. Morrison, ``String Theory on K3
Surfaces'', hep-th/9404151.}
\nref\Ac{P.S. Aspinwall, ``Enhanced Gauge Symmetries and K3
Surfaces'', Phys. Lett. B357 (1995) 329,  hep-th/9507012.}
\nref\S{ A. Strominger, ``Open P-Branes'', Phys. Lett. B383
(1996) 44, hep-th/9512059.}
\nref\T{ P.K. Townsend, ``Brane Surgery'', Proceedings of the
European Research Conference on 'Advanced Quantum Field Theory'. La
lonnde les Maures, France, Sept. 1996.}
\nref\Ir{N. Seiberg, ``IR Dynamics on Branes and Space-Time
Geometry'', Phys.Lett. B384 (1996) 81, hep-th/9606017.}
\nref\SW{ N. Seiberg, E. Witten, ``Gauge Dynamics And
Compactification To Three Dimensions'', hep-th/9607163.}
\nref\K{ R.Khuri, ``A Multimonopole Solution In String Theory'',
Phys. Lett. B294 (1992) 325, hep-th/9205051,\parskip=0pt
\item{}R.Khuri, ``A Heterotic Multimonopole Solution'', Nucl. Phys. B387
(1992) 315, hep-th/9205081,
\item{}R.Khuri, ``Monopoles and Instantons in String Theory'', Phys. Rev. D46
(1992) 4526, hep-th/9205091,
\item{}J.P.Gauntlett, J.A. Harvey, J.T. Liu, ``Magnetic Monopoles in String
Theory'', Nucl. Phys. B409 (1993) 363, hep-th/9211056.}
\nref\DVVb{R. Dijkgraaf, E. Verlinde, H. Verlinde, ``5D Black Holes
and Matrix Strings'', hep-th/9704018.}
\nref\qm{M. Claudson, M. B. Halpern, ``Supersymmetric ground state
wave function'', Nucl. Phys. B250 (1985) 689,\parskip=0pt 
\item{}R. Flume, ``On Quantum Mechanics with extended
supersymmetry and nonabelian gauge constraints'', 
Annals Phys.164 (1985) 189,
\item{}M. Baake, M. Reinicke, V. Rittenberg, ``Fierz identities for
real Clifford algebras and the number of supercharges'',
J.Math.Phys.26 (1985) 1070,\parskip=0pt 
\item{} V. Rittenberg, S. Yankielowicz, ``Supersymmetric gauge
theories in quantum mechanics'',  Annals Phys.162 (1985), 273.}
\nref\DM{ M.R. Douglas, G. Moore, `` D-branes, Quivers and ALE
Instantons'', hep-th/9603167.}
\nref\Kr{ P.B. Kronheimer, ``The construction of ALE spaces as
hyper-kahler quotients'', J. Diff. Geom. 28 (1989) 665.}
\nref\N{ H. Nakajima, ``Instantons on ALE spaces, quiver
varieties, and Kac-Moody algebras'', Duke Math. 76 (1994) 365.}
\nref\DGM{ M.R. Douglas, B.R. Greene, D.R.Morrison, ``Orbifold
Resolution by D-Branes'', hep-th/9704151.}
\nref\P{J. Polchinski, ``Tensors from K3 Orientifolds'', 
Phys.Rev. D55 (1997) 6423, hep-th/9606165.}
\nref\DDG{ D.-E. Diaconescu, M.R. Douglas, J. Gomis, to appear}
\Title{\vbox{\baselineskip12pt\hbox{hep-th/9707019}
\hbox{RU-97-53}}}
{\vbox{
\centerline{Duality In Matrix Theory And}
\vskip 10pt
\centerline{Three Dimensional Mirror Symmetry}
}}
\centerline{Duiliu-Emanuel Diaconescu and Jaume Gomis}
\medskip
\centerline{\it Department of Physics and Astronomy}
\centerline{\it Rutgers University }
\centerline{\it Piscataway, NJ 08855--0849}
\medskip
\centerline{\tt duiliu, jaume@physics.rutgers.edu}
\medskip
\bigskip
\noindent

Certain limits of the duality between M-theory on ${T^5/Z_2}$ and IIB
on K3 are  
analyzed in Matrix theory. The correspondence between M-theory
five-branes and ALE backgrounds is realized as
three dimensional mirror symmetry. 
Non-critical strings dual to open membranes  are explicitly 
described as gauge theory excitations. 
We also comment on Type IIA on K3
and the appearance of gauge symmetry enhancement at special points in
the moduli space.

\Date{July 1997}

\newsec{Introduction}

Recently, it has become clear that the non-perturbative
formulation of M-theory proposed by Banks,
Fischler, Shenker and Susskind \BFSS\ captures many essential aspects 
of conventional string physics. Various tests of this theory
include the derivation of elementary strings and their interactions
\refs{\Motl, \BS, \DVV},
construction and scattering of solitonic states \refs{\BFSS, \BD,
\BSS, \scatt}, 
compactifications on tori and  K3 surfaces 
\refs{\GRT, \D, \FHR, \FR, \KR, \FS, \BRS, \Seib, \gov, \BR},
and finally realizations of standard dualities \refs{\GRT, \gov, \BR}.

The purpose of the present paper is to study the Matrix theory
description of the $\hbox{M-theory on}\ T^5/Z_2\leftrightarrow
\hbox{IIB on}\ K3$ duality \refs{\Wa}. Since compactification of Matrix theory
on high dimensional tori involves certain subtleties \refs{\BRS,
\Seib} we restrict to 
special limits in the moduli space where both models can be realized 
as three dimensional gauge theories. In M-theory this corresponds to a
degeneration of the five-torus of the form $T^5\simeq T^2\times R^3$
while in IIB theory we consider non-compact 
K3 surfaces described by ALE gravitational instantons.
It turns out that this framework allows a realization of the 
conventional $\hbox{ALE}\leftrightarrow\hbox{five-brane}$ duality \VO\ 
as three dimensional mirror symmetry \refs{\IS, \OO, \HW, \PZ} as
summarized below.
\vskip 1pt
$\bullet$ A configuration of $n$ five-branes away from the fixed point
in $(T^2\times R^3)/Z_2$ is described by an $\scrn=4$ $U(k)$ gauge theory with 
$n$ fundamental hypermultiplets and one adjoint. The mirror pair is
defined by an $A_{n-1}$ quiver and describes IIB theory on an
$A_{n-1}$ ALE background. Non-critical Type IIB strings filling out an
$SO(n+1,2,Z)$ multiplet are
identified as excitations in the gauge theory. 
By mirror symmetry we obtain a gauge theory description of 
open membranes stretched between five-branes.
\vskip 1pt
$\bullet$ A configuration of $n$ five-branes near  the fixed point 
is described by an $\scrn=4$ $Sp(k)$ gauge theory with $n$ fundamental
hypermultiplets and one anti-symmetric tensor. The mirror pair is defined by 
an $D_n$ quiver and describes IIB theory on a $D_n$ ALE background.
Again, we identify the $SO(n+2,2,Z)$ multiplet of non-critical strings
providing a mirror description of dual open membranes.

We also consider the  Matrix realization of IIA theory on an ALE
space emphasizing the phenomenon of gauge symmetry enhancement in the
orbifold limit. The description of the relevant wrapped D2-brane
states parallels that of Type IIB non-critical strings.

Section $2$ is a brief review of certain aspects of the 
conventional theories introduced above. In section $3$
we describe five-brane backgrounds in Matrix theory compactified
on two tori. The realization of IIA and IIB models is developed 
in section $4$.
Section $5$ consists of a detailed presentation of the mirror map
between non-critical strings and open membranes.

\newsec{Type IIA and Type IIB on K3 and
M-theory on $T^5/Z_{2}$}

The moduli space of Type IIB theory on K3 is \refs{\Wc, \Ad, \Aa,
\HD} the Grassmanian of
space-like five-planes in $R^{21,5}$ modulo discrete identifications
\eqn\Bmoduli{
{\cal M}=SO(21,5,Z) \backslash SO(21,5)/{(SO(21)\times SO(5))}}
This can be equivalently regarded as the moduli space of even
self-dual lattices $\Gamma^{21,5}$ representing the K3 cohomology
lattice $\Gamma^{19,3}$ supplemented by two copies of the hyperbolic
plane $\Gamma^{1,1}\oplus \Gamma^{1,1}$ corresponding to $B$ and
$C$ moduli. The six-dimensional dynamics is characterized \refs{\Wd,\HD} by 
the occurrence of an $SO(21,5,Z)$ multiplet of non-critical BPS strings 
whose charges are vectors in $\Gamma^{21,5}$. The states represented
by charge vectors in $\Gamma^{19,3}\subset\Gamma^{21,5}$ can be 
identified as wrapped D3-brane states with a BPS tension
formula 
\eqn\BPSB{
T={1\over g_s}\sqrt{B^2+C^2+|\Omega|^2+J^2}}
where $\Omega$, $J$ are the values of the holomorphic two-form and
K\"ahler class on the cycle.

We will consider the theory locally near a blown-up ADE singularity
embedded in a K3 surface of very large radius. The cohomology lattice
spanned by the exceptional cycles is $\Gamma^{n-1,0}$ for $A_{n-1}$ 
ALE and $\Gamma^{n,0}$ for $D_n$ ALE. Accordingly, the non-critical
strings will fill out $SO(n+1,2,Z)$ and $SO(n+2,2,Z)$ multiplets 
respectively.

Type IIA theory has a similar moduli space \refs{\Wc, \Aa, \Ab, \Ac}
parameterized by the quotient
\eqn\Amoduli{
{\cal M}=SO(20,4,Z) \backslash SO(20,4)/{(SO(20)\times SO(4))}}
The six-dimensional theory contains in this case BPS particles rather
than strings, a subset of which can be identified with wrapped D2-brane
states. The BPS mass formula is analogous
\eqn\BPSA{
m={1\over g_s}\sqrt{B^2+|\Omega|^2+J^2}}

According to \refs{\Wa} Type IIB theory on K3 is dual to M-theory
on $T^5/Z_2$. Anomaly cancelation and magnetic charge conservation
requires 16 M-theory five-branes in the background. 
The non-critical strings are dual \refs{\Wa, \HD}
to open membranes stretching between
five-branes \refs{\S, \T}. In particular configurations
with multiple five-branes are dual to ADE singularities in the K3
surface. As stated in the introduction we consider a particular limit
of the moduli space in which the five-torus degenerates as 
$T^5\simeq T^2\times R^3$. This creates two distinct situations 
depending whether the five-branes coalesce at the fixed point or 
at a distant point in $R^3$. The relevant singularities are
respectively $A_{n-1}$ and $D_n$, where $n$ is the number of branes. 

\newsec{Matrix on Tori and Five-Branes}

It has been shown in \Seib\ that compactification of Matrix theory on 
$T^5/Z_2$ can be described in terms of a new six-dimensional theory
with eight supercharges modeled by heterotic five-branes at zero
coupling. The theories studied there are related to $Spin(32)$
instantons of charge $k$ but we will need a slight modification 
corresponding to $Spin(n)$ instantons. The low energy limit is a
six-dimensional $\scrn=(1,0)$ $Sp(k)$ gauge theory with $n$
fundamental hypermultiplets and one antisymmetric tensor. 
As emphasized in \refs{\BRS, \Seib}, compactification of the full
theory on a generic five-torus does not have a well defined moduli space.
However, in the case of interest here 
the space-time torus is $T^5\simeq T^2\times R^3$ so the dual torus 
degenerates to $\tilde T^2\times T^3$ with $T^3$ shrinking to zero size.
The low energy limit reduces to 
an $\scrn=4$ three-dimensional gauge theory on $R\times \tilde T^2$
with the same field content
describing a configuration of $n$ five-branes coming together at the
fixed point in $T^2\times R^3$. 
Note that 
the same effective description has been realized earlier in \refs{\KR,
\FS} starting from the original definition of Matrix theory
as quantum mechanics of D0-branes.
Also the construction outlined above incorporates automatically
the fundamental hypermultiplets which describe longitudinal five-branes \BD.

Moving away from the fixed point, the geometry is locally $T^2\times
R^3$ since the $Z_2$ projection relates tori at different points in
$R^3$. A configuration of $n$ coinciding five-branes will be described
by a three-dimensional gauge theory with $U(k)$ gauge group,
$n$ hypermultiplets in the fundamental and one in the adjoint. 
In the picture of \refs{\KR, \FS}, this can be understood as a 
splitting of the original configuration of
$2k$ D0-branes in two equal groups far apart from each other and decoupling
of heavy open strings that stretch from one group to the other.
The $Z_2$ projection simply relates the two separated groups
so the gauge group is actually $U(k)$.

Further evidence supporting this picture comes from identifying 
the Coulomb branches of the gauge theories with five-brane
backgrounds. We will outline the details only for $U(k)$ theories, the
$Sp(k)$ being analogous. 
The parameters of the gauge theory are \refs{\FHR}
\eqn\Bparameters{
{1\over g^2}={ L_1L_2\over (2\pi)^2R},\qquad
\Sigma_m= {(2\pi)^2l_{11}^3\over RL_m},\qquad m=1,\,2}
where $L_m$ are the circumferences of the space-time torus. 
As argued in \FHR,\ Type IIB is obtained in the limit $L_m\rightarrow
0$ with $L_2/L_1<<1$ fixed. In this regime the dual torus approaches
the decompactification limit at fixed $\Sigma_1/\Sigma_2<<1$.

The result can be viewed as either IIB theory on a circle of
circumference
\eqn\LY{
L_Y=(2\pi)^3{l_{11}^3\over L_1L_2}}
or, by T-duality, IIA theory on a circle of circumference $L_2$ \FHR.
The string scale and the Type IIB coupling constant are given by
\eqn\alph{\alpha'={2\pi l_{11}^3\over L_2},\qquad
g_s={L_2\over L_1}.}

After dualizing the low energy photons, the $U(k)$
reduces essentially to $k$ identical copies of a $U(1)$ gauge theory
with the same field content. Thus it suffices to study the Coulomb
branch of the latter. In vector multiplet variables $(\vec\phi,
A_\mu)$ the one-loop corrections are exact \refs{\Ir, \SW} and consist of
the metric 
\eqn\metric{
ds^2=\left({1\over g^2}+\sum_{i=0}^{n-1}{1\over |\vec \phi-\vec
m_i|}\right)d\vec \phi\cdot d\vec \phi.}
and a Chern-Simons term
\footnote{${}^{1}$}{Actually this term is obtained from that of
\refs{\Ir, \SW} by an integration by parts.}
\eqn\CS{
\sum_{i=0}^{n-1}\epsilon^{\lambda\mu\nu}\vec\omega_{i}\cdot\partial_\lambda
\vec \phi F_{\mu\nu}}
where $\omega_i$ are standard Dirac monopole vector potentials
and $\vec m_i$ are hypermultiplet bare masses. 
Since the dual torus $\tilde T^2$ grows as $R\times S^1$ 
the theory is effectively compactified on a circle of very large
radius. Dimensional reduction of the three dimensional Coulomb branch 
\footnote{${}^{2}$}
{Strictly speaking, this procedure is valid only far
out on flat directions where the radius of the circle is much larger 
than the length scale $1\over |\vec \phi|$ and it should break down 
in a central region of the moduli space. However, this region can be made
arbitrarily small by making the radius of the circle arbitrarily big. 
This is the regime in which the three dimensional description is
valid.}
yields an $\scrn=(4,4)$ two dimensional sigma-model with torsion
\eqn\sigmamodel{
{1\over 2\pi\alpha'}\int\,d\sigma_2 d\tau
\left\{\left(1+\sum_{i=0}^{n-1}{2\pi\alpha'\over 
R_1}{1\over |\vec r-\vec r_i|}\right)\left({d\vec r}^2+R_1^2
d\theta_1^2\right)+2\pi\alpha'\sum_{i=0}^{n-1}\epsilon^{\mu\nu}\partial_\mu\vec
\omega_i\cdot  
\partial_\nu\vec r\right\}.}
where the conversion between space-time and gauge theory parameters is
given by
\eqn\conv{
\vec r= {2\pi l_{11}^3\over R_{2}}\vec \phi,\qquad 
\vec r_i={2\pi l_{11}^3\over R_{2}}\vec m_i.}
It is easy to check that this is exactly the symmetric monopole
solution of \refs{\K}, thus the gauge theory indeed describes
five-brane backgrounds as claimed before. A similar check
has been performed in two dimensional context in \DVVb.

\newsec{Matrix description of string theory on ALE spaces}

The standard approach to the IMF definition of M-theory 
on non-compact
ALE backgrounds \refs{\D, \FR} is the quantum mechanics
\footnote{${}^3$}{This quantum mechanics was first considered in \qm.}
of D0-branes on a
$C^2/\Gamma$ orbifold \refs{\DM}. However it has been argued 
recently, \refs{\BRS, \Seib, \BR} that a complete treatment
of compact K3 surfaces involves new six-dimensional theories
compactified on the dual K3 times a circle. In particular 
certain degrees of freedom related to the exceptional cycles 
of a blown-up singularity are absent from the usual construction. 
Since at the present stage it is not clear how to make the latter
description explicit, we will consider the first approach. 
Even if this should be regarded as an approximate description of the 
theory, it will turn out  that it 
captures the description of wrapped D-brane states.

\subsec{IIA on ALE}

As argued in \refs{\D, \FR} this theory can be realized in terms of a 
two dimensional $\scrn=(4,4)$ gauge theory on $S^1\times R$ 
with field content specified by a quiver diagram. 
The parameters of the gauge theory have been determined in \refs{\FHR}
\eqn\IIA{
 {1\over g^2}={\Sigma L^2\over (2\pi)^2R},\qquad  
 \Sigma L= {(2\pi)^2l_{11}^3\over R}.}
In addition we introduce Fayet-Iliopoulos parameters which can
be related to the blow-up modes present in the original quantum
mechanics by
\eqn\conversion{
\vec\zeta_{gauge}={R^2\over {(2\pi)^2l_{11}^6}}\vec\zeta_{string}.}
This is just the usual conversion between length in space-time and mass in
gauge theory. From now on the gauge theory parameters will be denoted simply
by $\vec\zeta$.
\vskip 2pt
$\bullet\ A_{n-1}\ ALE.$
The relevant gauge dynamics is related to  the Higgs branch
described by a hyper-K\"ahler quotient construction \refs{\Kr, \N}
\eqn\higgs{\eqalign{
\mu^R_0 \equiv &\ 
X_{01}X_{01}^{\dagger}-X_{10}^{\dagger}X_{10}+X_{0,n-1}X_{0,n-1}^{\dagger}
-X_{n-1,0}^{\dagger}X_{n-1,0}={\zeta^R_0}\,I_{k\times k} \cr
\mu^C_0 \equiv &\  X_{01}X_{10}-X_{0,n-1}X_{n-1,0}
=\zeta^C_0\,I_{k\times k} \cr
& \vdots \cr
\mu^R_{n-1} \equiv &\ 
X_{n-1,0}X_{n-1,0}^{\dagger}-X_{0,n-1}^{\dagger}X_{0,n-1}+X_{n-1,n-2}
X_{n-1,n-2}^{\dagger} \cr
 &\ -X_{n-2,n-1}^{\dagger}X_{n-2,n-1}
={\zeta^R_{n-1}}\,I_{k\times k}\cr
\mu^C_{n-1} \equiv &\ 
X_{n-1,0}X_{0,n-1}-X_{n-1,n-2}X_{n-2,n-1}=\zeta^C_{n-1}\,I_{k\times
k}\cr}}
where the $X$'s are in one-to-one correspondence with the links of the
following quiver diagram

\ifig\quiverA{$A_{n-1}$ quiver diagram.}
{\epsfxsize1.5in\epsfbox{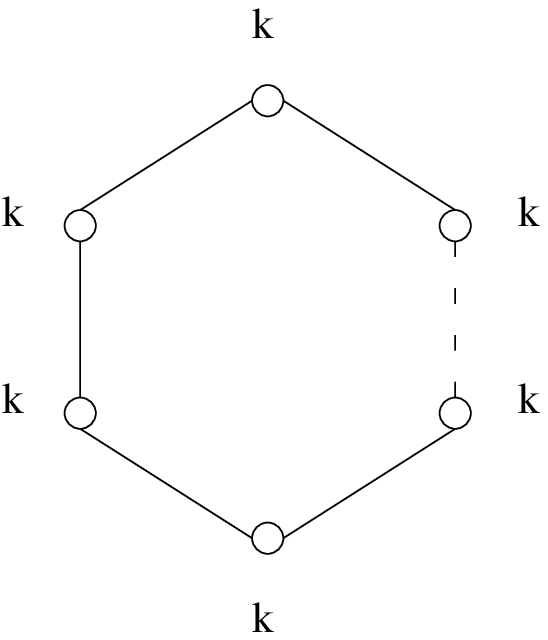}}

\noindent 
and the Fayet-Iliopoulos parameters are constrained by
\eqn\FIconstr{
\sum_{i=0}^{n-1}\vec\zeta_i=0.}
The moduli space of solutions modulo gauge transformations is
isomorphic to the symmetric product 
$Sym^k(\widetilde X_{A_{n-1}})$ of resolved ALE varieties of $A_{n-1}$ type. 
There is no Coulomb branch and the gauge group is generically broken
to the diagonal torus $U(1)_D^k\subset U(k)^n$ along the
hypermultiplet flat directions. 
Note that there is a one to one correspondence between vertices of the
extended Dynkin diagram and homology two cycles $\gamma_i$ of the 
blown-up ALE satisfying the linear constraint
\eqn\cycleconstr{
\sum_{i=0}^{n-1}\gamma_i=0.}

The schematic form of the action obtained by reduction from six
dimensions reads
\eqn\IIAact{
{\eqalign{
{1\over g^2}\int\, dt d\sigma\,\Bigl\{{1\over 2n}
&Tr\Bigl(F_0^2+F_1^2+\ldots +F_{n-1}^2
+{\vec D_0}^2+{\vec D_1}^2+\ldots +{\vec D_{n-1}}^2\Bigr)+\cr
&Tr\Bigl(\vec D_0\cdot
(\vec\mu_0-\vec\zeta_0\,I_{k\times k})
+\vec D_1\cdot(\vec\mu_1-\vec\zeta_1\,I_{k\times k})+\ldots \cr
&\vec D_{n-1}\cdot 
(\vec\mu_{n-1}-\vec\zeta_{n-1}\,I_{k\times k})\Bigr)\Bigr\}\cr}}}
where $\vec D$ are triplets of auxiliary fields in the
six-dimensional vector multiplet and $\vec \mu$ are the moment maps
introduced in \higgs. The factor ${1/n}$ multiplying the gauge kinetic
terms is inherited from the original quantum mechanics where
a single D0-brane moving on the ALE
is described by $n$ identical images under the orbifold
group $Z_n$. Thus the normalization is fixed such that summing over all
images yields the correct kinetic term. 

Different states in the resulting Type IIA theory can be identified
with gauge theory excitations as follows. D0-branes propagating freely
on the ALE background can be described by turning on t'Hooft fluxes 
${1\over kn}I_{k\times k}$ in each of the $n$ factors of the overall 
diagonal $U(1)\subset U(k)^n$. The energy of a single 
quantum of flux is computed by summing over all $n$
fluxes and the correct scaling behavior is ensured by the
factor $1/n$ mentioned above 
\footnote{${}^{4}$}{We thank T. Banks for  clarifying
explanations on this point.}
\eqn\fluxA{
E_E={R\over 2k}\left({2\pi\over L}\right)^2={1\over 2p_{11}}
\left({1\over g_s\sqrt\alpha'}\right)^2.}
It matches the light cone energy of a single D0-brane carrying
$k$ units of longitudinal momentum. 
The string theory parameters
present in this formula characterize the newly defined theory and are
given by
\eqn\stringpar{
\alpha'={2\pi l_{11}^3\over L},
\qquad
g_s^2={1\over (2\pi)^3}{L^3\over l_{11}^3}.}

As stated in section one the theory is expected to contain more
general BPS states with charge vectors in the cohomology lattice
$\Gamma^{n-1,0}$. In conventional string theory these have 
been identified as fractional D0-branes described by a single image 
of the orbifold gauge group \refs{\D, \DGM, \P}. 
The hypermultiplet degrees of freedom corresponding to 
motion on the ALE are projected out. Therefore the object is stuck
at the fixed point but free to move in transverse directions.

In Matrix gauge theory these states can be described as positive
energy excitations of the hypermultiplets away from the Higgs branch
\D. In order to identify wrapped states on the $i$th
cycle $\gamma_i$ we propose to set
the expectation values of the hypermultiplets charged under $U(k)_i$
to zero \footnote{${}^{5}$}{We thank M. Douglas for suggesting
this.}. We will refer to this mechanism as  
amputation of the quiver diagram (see fig. 2).
This results in a positive energy field configuration sitting at a 
stationary point of the potential. 
The scalars in the $U(k)_i$ vector multiplet will
parameterize a new ``Coulomb'' branch of 
{\it meta-stable flat directions}. 

\ifig\quiverB{Amputation of adjacent legs in an $A_{n-1}$ quiver
diagram. This describes a wrapped state on the isolated cycle.}
{\epsfxsize1.5in\epsfbox{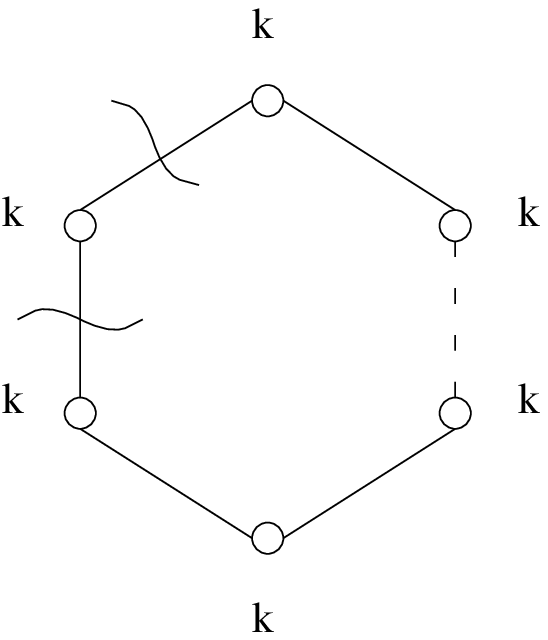}}

To prove that the energy of these excitations has the correct scaling
behavior, note that the gauge group is $U(k)=(SU(k)\times
U(1))/Z_k$ and that we have matter in fundamental representation which 
is acted faithfully by the center $Z_k$. Therefore by 
normalizing the $SU(k)$ charges to be integers, the $U(1)$ charges
have to be quantized in $1/k$ 
\footnote{${}^6$}{We thank N. Seiberg and S. Shenker for
explaining this to us.}. This factor will affect all couplings
of the diagonal $U(1)\subset U(k)$ with fundamental hypermultiplets.
Integrating out the diagonal $\vec D$ terms in \IIAact\
results in an energy 
\eqn\energyA{
\Delta E={1\over 2}{nR\over k}{1\over
{(2\pi)^2l_{11}^6}}\{|\vec\zeta_i|^2+{1\over n-1}|\sum_{j\neq i}
\vec\zeta_j|^2\}}
In the IMF this corresponds to an overlap of two states with mass and
longitudinal momentum given by 
\eqn\massA{\eqalign{
& M^2={|\vec\zeta_i|^2\over
{(2\pi)^2g_s^2 \alpha'^3}}\qquad 
p_{11}={1\over n}{k\over R}\hfill\cr
& {M^\prime}^2={|\sum_{j\neq i}\vec\zeta_j|^2\over
{(2\pi)^2g_s^2 \alpha'^3}}\qquad {p^\prime}_{11}=
{n-1\over n}{k\over R}\cr}.}
Note that \FIconstr\ implies that $M=M^\prime$.
We interpret this as a configuration of two D2-branes 
with opposite charges wrapped on the
cycle $\gamma_i$ and the complementary cycle $\sum_{j\neq
i}\gamma_j=-\gamma_i$.
Therefore these states can be viewed as individual stable particles
only if they are 
separated far apart in transverse directions. In principle they can 
come together and annihilate into uncharged states
\footnote{${}^7$}
{We thank M. Douglas for ample explanations on these issues.}.
A detailed study of these issues will appear elsewhere \DDG.
The B-field on cycles is zero, thus these
states become massless in the orbifold limit. This should be
contrasted with the situation in the original IIA theory \refs{\D,
\DGM}. 

\ifig\quiverC{Amputation of non-adjacent legs in an $A_{n-1}$ quiver
diagram. This describes a bound state of wrapped states on the two
isolated cycles.}
{\epsfxsize1.5in\epsfbox{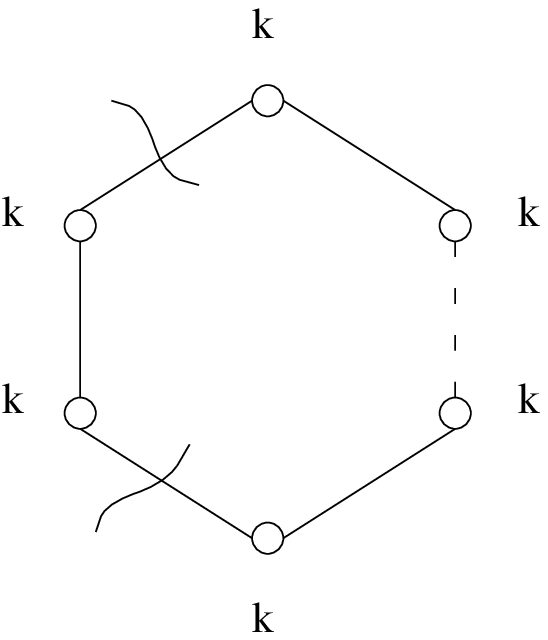}}

This procedure can be generalized to obtain bound states 
of D2-branes wrapped on linear combination of cycles 
$\sum_{i=0}^{p-1}\gamma_i$. The links of the quiver diagram 
are cut off such that one is left with two separate 
groups of $p$ and respectively $n-p$ interconnected vertices. 
The case $p=2$ is represented in \quiverC. In this case the 
energy is 
\eqn\energyB{
\Delta E={1\over 2}{nR\over k}{1\over {(2\pi)^2l_{11}^6}}
\left\{{1\over p}|\sum_{i=0}^{p-1}\vec\zeta_i|^2 +
{1\over n-p}|\sum_{i=p}^n\vec\zeta_i|^2\right\}.}
As before, we interpret this configuration as two wrapped states with 
opposite charges. The masses and longitudinal momenta read
\eqn\massB{\eqalign{
& M^2={|\sum_{i=0}^{p-1}\vec\zeta_i|^2\over
{(2\pi)^2g_s^2 \alpha'^3}}\qquad\ p_{11}={p\over n}{k\over R}\cr
& {M^\prime}^2={|\sum_{i=p}^n\vec\zeta_i|^2\over
{(2\pi)^2g_s^2 \alpha'^3}}\qquad p^\prime_{11}
={(n-p)\over n}{k\over R}\cr}.}
Therefore we obtain bound states in $1-1$ correspondence with the
roots of the Dynkin diagram filling out  
the expected non-Abelian vector multiplet in the orbifold limit.
\vskip 2pt
$\bullet\ D_n\ ALE.$
These models can be described similarly to the $A_{n-1}$ case with
minor differences. 
The gauge theory is defined by the regular $D_n$ quiver diagram.
The hyper-K\"ahler moment maps defining the Higgs branch 
$Sym^k(\tilde X_{D_n})$ are similar to \higgs. The action is
essentially obtained from \IIAact\ by replacing the factor $n$ 
with the dual Coxeter number $c_2=2(n-1)$.

String theory states can be described similarly to the $A_{n-1}$
case. D0-branes propagating freely on the ALE are identified with 
t'Hooft fluxes ${1\over c_2k}$ in the overall diagonal 
$U(1)\subset U(k)^4\times U(2k)^{n-3}$. 

\ifig\quiverD{$D_n$ quiver diagram.}
{\epsfxsize3.5in\epsfbox{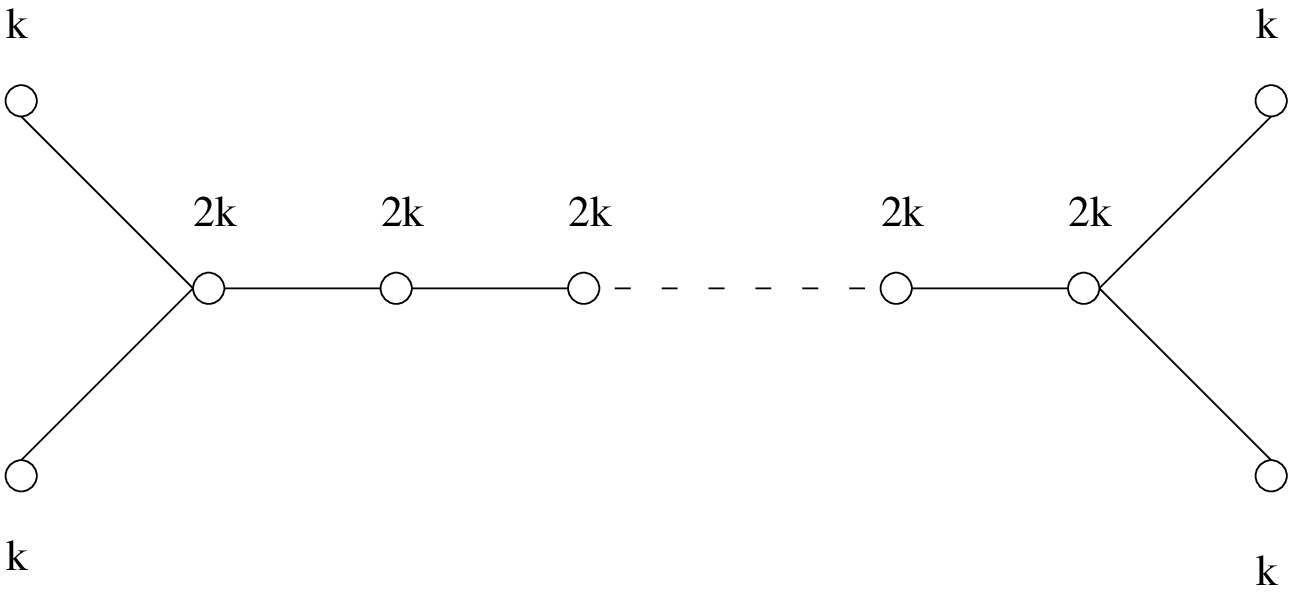}}

\noindent Excitations corresponding to
wrapped D2-branes can be obtained by exciting the hypermultiplets 
as before. In this case we must distinguish between cycles represented 
by outer and respectively inner vertices of the extended Dynkin
diagram. 
Amputation of the $D_n$ quiver will generally result 
in configurations of several wrapped states whose charges sum to zero.
The IMF mass and longitudinal momenta can be computed similarly to
\energyA. 

\subsec{IIB on ALE}

Type IIB on an $S^1\times {\rm{ALE}}$ background can be constructed 
as the three dimensional counter-part of the quiver gauge theories 
studied in the previous section. More precisely compactification
of the orbifold quantum mechanics on a two torus yields 
the three dimensional $\scrn=4$ quiver gauge theories studied 
extensively in \refs{\IS, \OO, \HW, \PZ}.
The parameters of the gauge theory as well as those of the derived
string theory have been specified in \Bparameters-\alph\ and \conversion.
We consider again two cases:
\vskip 2pt
$\bullet\ A_{n-1}\ ALE.$ The model is defined by the quiver in
\quiverA.
The gauge dynamics on the Higgs branch is identical to the two
dimensional theory but the string interpretation of 
the excitations is different. Electric fluxes in the overall
diagonal $U(1)\subset U(k)^n$ can be identified
with IIB fundamental or D-strings propagating on the ALE background.
The energy of a single quanta of flux in direction $m=1,2$ is:
\eqn\fluxB{
E_E={R\over 2k}\left({L_YL_m\over (2\pi)^2 g_{s}\alpha '^{3/2}}\right)^2.}
For $m=1$ this matches the IMF energy of a D-string wrapped on $S^1_Y$
while for $m=2$ this matches the IMF energy of a wrapped fundamental
string as in \FHR. Again the fact that these strings are free to
propagate on the ALE is consistent with the diagonal $U(1)$ being 
generically unbroken on the Higgs branch.

As explained in section one IIB theory on an ALE space has in addition
an $SO(2,n+1,Z)$ multiplet of non-critical BPS strings whose charge
vectors live in the lattice $\Gamma^{0,n-1}\oplus\Gamma^{2,2}$. The
first factor is the homology lattice of the resolved ALE space and
the corresponding non-critical strings are actually D3-branes wrapped
on the homology two cycles. These are T-dual to wrapped 
D2-branes in Type IIA theory, thus we know how to identify them
as meta-stable positive energy excitations in the gauge theory.
The mechanism is the same and 
we only need to check that the $\vec D$ term energy has 
the correct scaling behavior. Amputation of the quiver diagram as in 
\quiverB\ results in a configuration with energy
\eqn\energyC{
\Delta E=
{nR\over 2k}{1\over (2\pi)^2 l_{11}^6}\left
\{|\vec\zeta_i|^2+{1\over n-1}|\sum_{j\neq i}
\vec\zeta_j|^2\right\}}
In the IMF this corresponds to an overlap of two states with masses
and longitudinal momenta given by
\eqn\massC{\eqalign{
& M^2={|\vec\zeta_i|^2L_Y^2\over{(2\pi)^4 g_s^2 \alpha'^4}}
\qquad\quad\ \ p_{11}={1\over n}{k\over R}\cr
& {M^\prime}^2={|\sum_{j\neq i}\vec\zeta_j|^2
L_Y^2\over{(2\pi)^4 g_s^2 \alpha'^4}}\qquad 
{p^\prime}_{11}={n-1\over n}{k\over R}\cr}.}
As before, we interpret this as a metastable configuration of
D3-branes with opposite charges 
wrapped on the cycle $\gamma_i$ and the extra circle 
$S^1_Y$. In order to identify individual stable states they must 
be given a large separation in the transverse directions.

We can also identify non-critical strings corresponding to non-simple
roots in the homology lattice of the ALE as bound states of D3-branes
wrapped on linear combinations of cycles. The correct scaling behavior can be
shown in  a calculation identical to \energyB.
However, there are 
more general non-critical strings whose charge vectors have non-zero
components in the second factor $\Gamma^{2,2}$ that  carry charge
$(p,q)$ under the bulk two form potentials $B,\ C$. In conventional string
theory these may be identified as non-marginal
bound states of wrapped D3-branes
and $(p,q)$ strings whose mass is given by:
\eqn\boundmass{
M^2=T_3^2|\vec\zeta_i|^2L_Y^2+T_{(p,q)}^2L_Y^2.}
Realization of these states in Matrix picture is non-trivial
and involves crucially the new branch of meta-stable flat directions
associated to the amputated quiver diagram in \quiverB. Since the expectation
values of the hypermultiplets charged under $U(k)_i$ have been set to
zero, the full $U(k)_i$ is un-higgsed at the origin. 
The meta-stable flat directions are parameterized by scalars in the
vector multiplet thus there will be generically $k$ low energy
mass-less photons $U(1)^k$. We stress that these are not the same 
as the generic mass-less photons on the Higgs branch.
 
Turning on a t'Hooft flux ${1\over k}I_{k\times k}$
in the diagonal $U(1)\subset U(k)_i$ along $\Sigma_m$ 
increases the energy of an individual wrapped state on the
cycle $\gamma_i$ to
\eqn\energyD{\Delta E={nR\over 2k}
{|\vec\zeta_i|^2L_Y^2\over{(2\pi)^8g_s^2 \alpha'^4}}+
{nR\over 2k}\left({L_YL_m\over (2\pi)^2 g_{s}
\alpha '^{3/2}}\right)^2}
which results in an IMF mass
\eqn\massD{
M^2={|\vec\zeta_i|^2L_Y^2\over{(2\pi)^8g_s^2
\alpha'^4}}+\left({L_YL_m\over (2\pi)^2 g_{s}
\alpha^{3/2}}\right)^2.}
Taking into account \fluxB\ and \energyC,\ it is clear that this
result is in agreement with the one expected from string
theory. 
It is interesting to trace the origin of the factor of $n$ in the
energy of the flux quanta. This follows from the fact that in the
original quantum mechanics the gauge kinetic term for a single image 
is ${1\over 2ng_s}F^2$ as opposed to the gauge kinetic term for the
diagonal $U(1)$. 
We conclude that all non-critical strings in the $SO(2,n+1,Z)$
multiplet can be described in Matrix theory.
\vskip 2pt 
$\bullet\ D_n\ ALE.$ The analysis is very similar to the $A_{n-1}$
case and it will not be repeated here. As in the previous section we
must differentiate between outer and inner cycles. The states 
wrapped on inner cycles carry $2k\over c_2$ units of longitudinal
momentum while those wrapped on outer cycles carry $k\over c_2$ units.

\newsec{Duality and Mirror Symmetry}

\noindent The results obtained in the previous sections can be best summarized 
in the following diagrams
\smallskip
$$
{\matrix{
\hbox{Type IIB on}\ A_{n-1}\ \hbox{ ALE}\hfill&\longleftrightarrow &
\hbox{M-theory on}\ T^2\times R^3\ \hbox{with {\it n} five-branes}\hfill\cr
\wr & & \wr \cr
3D\ A_{n-1}\ \hbox{quiver gauge theory}& \longleftrightarrow & 
3D\ U(k)\ \hbox{gauge theory with {\it n} hypermultiplets}\cr}}
$$
\medskip
$$
{\matrix{
\hbox{Type IIB on}\ D_n\ \hbox{ ALE}\hfill&\longleftrightarrow &
\hbox{M-theory on}\ (T^2\times R^3)/Z_2\ \hbox{with {\it n}
five-branes}\hfill\cr
\wr & & \wr \cr
3D\ D_n\ \hbox{quiver gauge theory}& \longleftrightarrow & 
3D\ Sp(k)\ \hbox{gauge theory with {\it n} hypermultiplets}\cr}}
$$
\medskip
\noindent
The upper arrow denotes conventional duality, $\wr$ stands for 
Matrix theory realization while the lower arrow is at this stage
the missing link in the chain. It turns out that the diagram is
in fact closed by the {\it three dimensional mirror symmetry}
discovered recently in \refs{\IS} and studied further in \refs{\OO,
\HW, \PZ}. The pairs of gauge theories describing the above Matrix model
backgrounds are related by electric-magnetic duality which maps 
in a precise manner the Higgs/Coulomb branch of the first theory to the
Coulomb/Higgs branch of the second. The FI parameters of the
quiver theories are mapped to bare hypermultiplet masses.
Taking into account the geometric interpretation of the two sets of
parameters we have derived a Matrix theory realization of the
duality between coalescing five-branes and ADE singularities. 

Moreover, in M-theory the non-critical strings are membranes stretching
between five-branes which become tensionless as the five-branes approach
each other. Therefore we have also obtained a Matrix-theory
description of open membranes, although from a dual point of view.
Pictorially, the correspondence can be represented as follows
in the $A_{n-1}$ case.

\ifig\fivebA{The membrane configuration dual to an $A_{n-1}$ quiver.}
{\epsfxsize3.0in\epsfbox{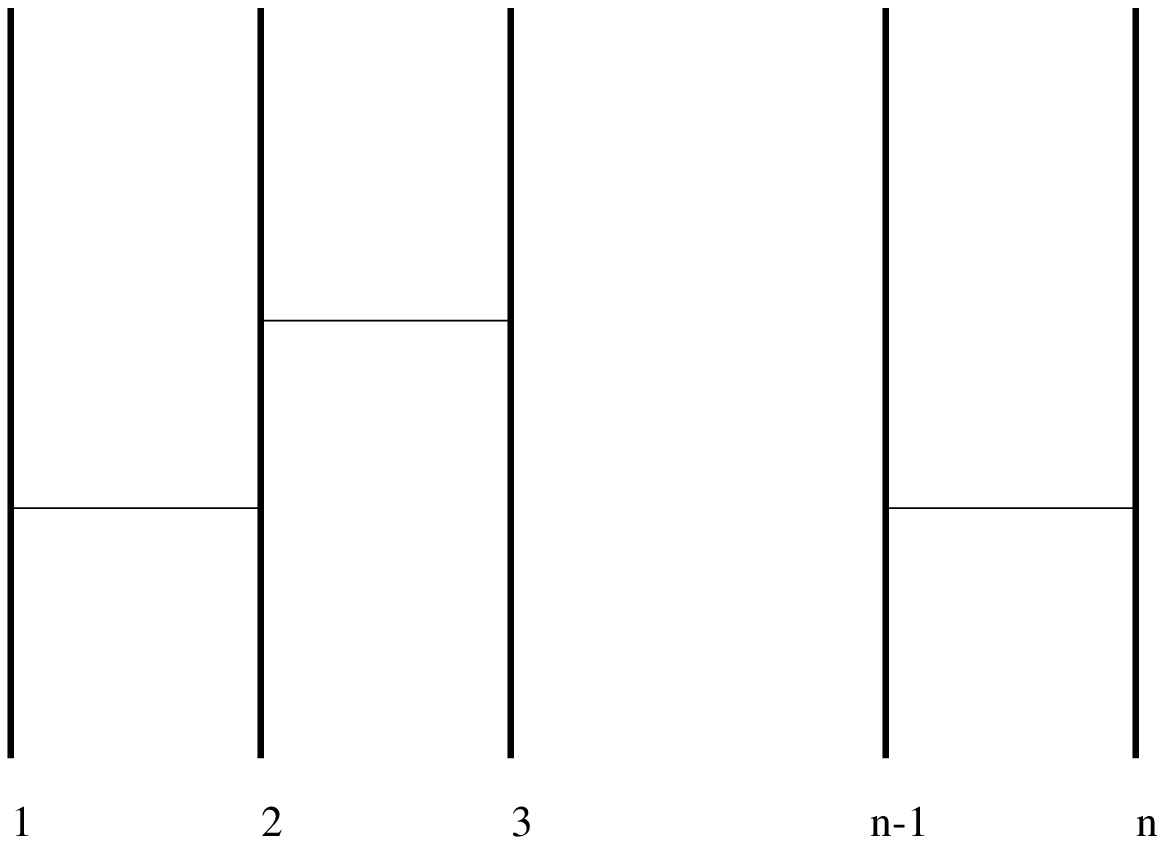}}

The simple roots of an $A_{n-1}$ Dynkin diagram can be represented in
terms of an orthonormal basis $\vec e_i$ in an $n$ dimensional vector
space as:
$$
\vec \alpha_1=\vec e_1-\vec e_2,\ \ldots\, 
,\vec \alpha_{n-1}=\vec e_{n-1}-\vec e_n.
$$
We can assign a five-brane to each vector such that a pair of consecutive
five-branes determines a simple root. Open membranes stretching between
five-branes are represented by horizontal line segments. The positions of
the five-branes are related to the sizes of the exceptional cycles by the
mirror map \refs{\IS, \OO}:
$$
\vec m_i=\sum_{l=0}^{i}\vec\zeta_l
$$
such that the tensions of open membranes match the masses of 
wrapped D3-branes.

\ifig\fivebB{The membrane configuration dual to a $D_n$ quiver.}
{\epsfxsize3.5in\epsfbox{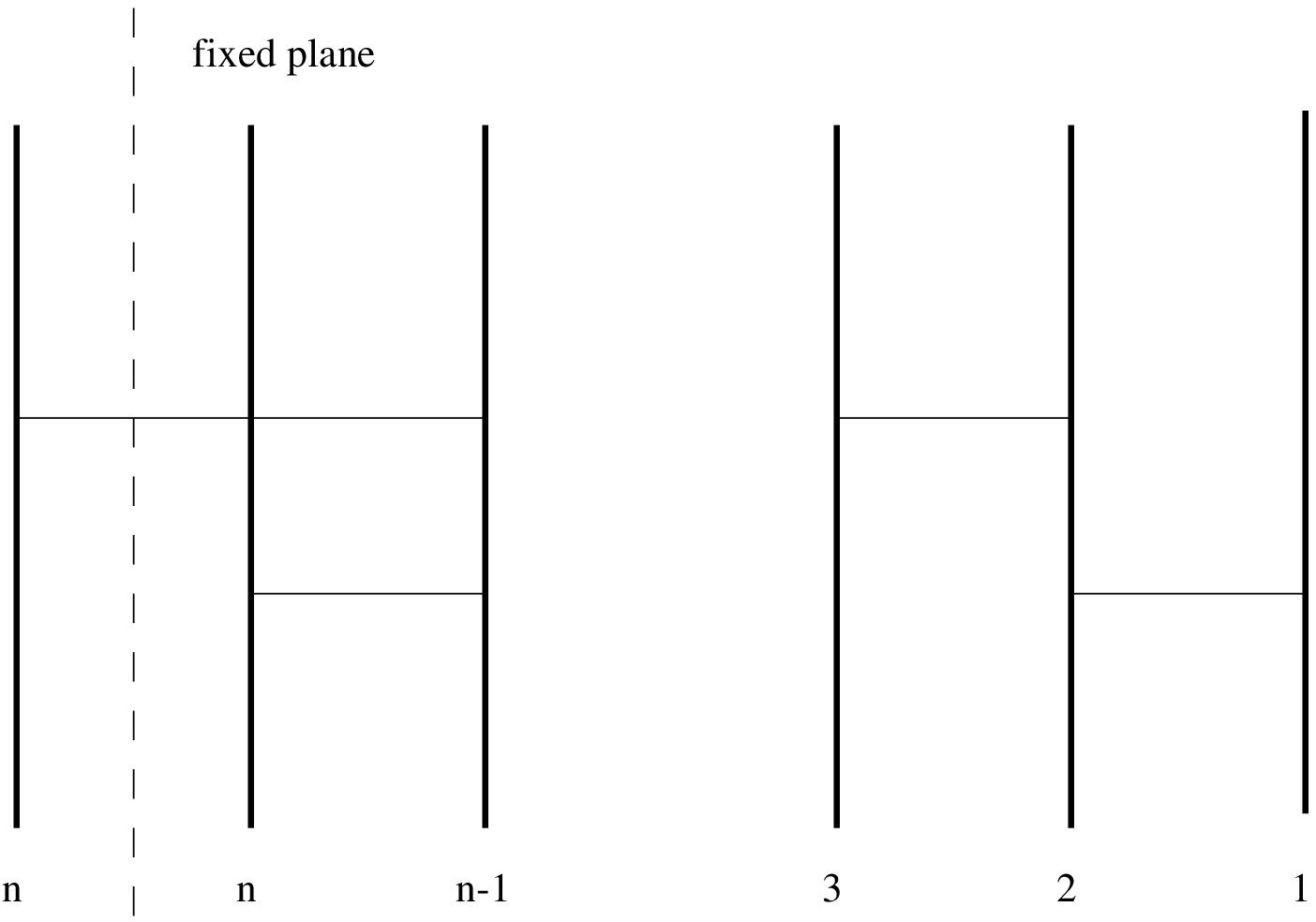}}

Similar considerations apply in the $D_n$ case.
The simple roots of a $D_n$ Dynkin diagram can be represented in terms
of the basis vectors as:
$$
\vec\beta_1=\vec e_1-\vec e_2,\ \ldots\, 
,\vec\beta_{n-1}=\vec e_{n-1}-\vec e_n,\ \vec\beta_n=\vec e_{n-1}+\vec
e_n.
$$
Note that the $n$th root is the reflection of the $(n-1)$th in the
hyperplane ${\vec e_n}^{\perp}$. When assigning as before a five-brane to
each basis
vector the last root will be determined by the $(n-1)$th five-brane and
the $Z_2$ image of the $n$th one. The mirror map is in this case:
$$
\vec m_n=\vec\zeta_n-\vec\zeta_{n-1},
$$
$$
\vec m_{n-1}=\vec\zeta_n+\vec\zeta_{n-1},
$$
$$
\vec m_{n-2}=2\vec\zeta_{n-2}+\vec\zeta_n+\vec\zeta_{n-1},
$$
$$
\vdots
$$
$$
\vec m_1=2\vec\zeta_1+2\vec\zeta_2+\ldots +2\vec\zeta_{n-2}
+\vec\zeta_n+\vec\zeta_{n-1}.
$$
D3-branes wrapped on the $i$th cycle  correspond to open membranes 
stretched between the $(i,\, i+1)$ five-branes except for $i=n$ when they
correspond to membranes stretched between the $(n-1)$th five-brane and
the $Z_2$ image of the $n$th five-brane. 
 
\newsec{Discussion}

In this paper, further evidence for the validity of Matrix theory as
the IMF description of eleven dimensional M-theory is presented. 
Certain local aspects of the 
conventional duality $\hbox{M}/(T^5/Z_{2}) \sim {\hbox{IIB}}/K3$ are 
realized in Matrix theory in terms of three dimensional mirror
symmetry. 

We also analyze ALE backgrounds in the Matrix theory definition of
Type IIA and Type IIB string theory. The expected multiplets of 
wrapped states are identified as gauge theory
excitations. The analysis reveals their bound state structure
in precise agreement with conventional string theory. Moreover
it is shown that mirror symmetry maps the description of non-critical 
IIB strings to open membranes stretching between M-theory five-branes. 

The SYM description used extensively above is only 
valid in special regimes. A full analysis would require 
detailed information on the six dimensional theories 
introduced in \refs{\BRS, \Seib}.
In section 3 we argued that the gauge theories
describing five-brane backgrounds can be regarded
as low energy limits of the full six dimensional $(1,0)$ theory
based on $Spin(32)$ instantons.
It would be very interesting 
to repeat the analysis for quiver gauge theories and the $(2,0)$
six dimensional string theory on $K3\times S^1$.
\bigskip
\centerline{\bf Acknowledgments}
We are indebted to Ofer Aharony, Micha Berkooz, Tom Banks, 
Michael Douglas, Nathan Seiberg and Steve Shenker
for very useful discussions and encouragement.

\vfill
\listrefs

\end